\documentclass[10pt,epsf]{article}
\usepackage{graphicx,amssymb}
\newcommand{\beq}{\begin{equation}}
\newcommand{\eeq}{\end{equation}}
\newcommand{\bsigma}{\mbox{\boldmath $\sigma$}}

\newcommand{\half}{\frac{1}{2}}
\newcommand{\La}{$\Lambda\,$}

\newcommand{\br}{{\bf r}}
\newcommand{\bqu}{{\bf q}}

\newcommand{\vbar}{{\overline V}}
\newcommand{\car}{$^{12}$C~}
\newcommand{\carla}{$^{12}_\Lambda$C~}

\newcommand{\oxy}{$^{16}$O }
\newcommand{\oxyla}{$^{16}_\Lambda$O~}

\newcommand{\ca}{$^{40}$Ca~}
\newcommand{\cala}{$^{40}_\Lambda$Ca~}

\newcommand{\zr}{$^{90}$Zr~}
\newcommand{\zrla}{$^{90}_\Lambda$Zr~}
\newcommand{\pb}{$^{208}$Pb~}
\newcommand{\pbla}{$^{208}_\Lambda$Pb~}

\newcommand{\threej}[6]{ \left( \begin{array}{ccc}
                               #1 & #2 & #3 \\
                               #4 & #5 & #6 
                             \end{array}
                        \right) } 
\newcommand{\sixj}[6]{ \left\{ \begin{array}{ccc}
                               #1 & #2 & #3 \\
                                #4 & #5 & #6 
                               \end{array}
                        \right\} } 
\newcommand{\ninej}[9]{ \left\{ \begin{array}{ccc}
                               #1 & #2 & #3 \\
                                #4 & #5 & #6 \\ 
                                #7 & #8 & #9 
                                \end{array}
                        \right\} } 
\def\ap{a_p}
\def\ap'{a_p'}
\def\ap1{a_{p_1}}
\def\ap'1{a_{p'_1}}
\def\ap2{a_{p_2}}
\def\ap'2{a_{p'_2}}
\def\ah{a_h}
\def\ah'{a_h'}
\def\ah1{a_{h_1}}
\def\ah'1{a_{h'_1}}
\def\ah2{a_{h_2}}
\def\ah'1{a_{h'_2}}
\def\acp{a\dag_p}
\def\acp'{a\dag_p'}
\def\acp1{a\dag_{p_1}}
\def\acp'1{a\dag_{p'_1}}
\def\acp2{a\dag_{p_2}}
\def\acp'2{a\dag_{p'_2}}
\def\ach{a\dag_h}
\def\ach'{a\dag_h'}
\def\ach1{a\dag_{h_1}}
\def\ach'1{a\dag_{h'_1}}
\def\ach2{a\dag_{h_2}}
\def\ach'1{a\dag_{h'_2}}
\begin{document}
%
%\begin{frontmatter}
%
\begin{center}
{\LARGE \bf A particle-hole model approach for hypernuclei}
\end{center}
\begin{center}
\vskip 0.5 cm 
{\large
M. Martini$^{\rm\,a,b}$, V. De Donno$^{\rm \,a}$, 
C. Maieron$^{\rm \, a}$, G. Co'$^{\rm \,\,\,a}$ \\
}
\vskip 0.5 cm 
$^{\rm a}$ Dipartimento di Fisica, Universit\`a del Salento, \\
and Istituto Nazionale di Fisica Nucleare  sez. di Lecce, \\
I-73100 Lecce, Italy \\
\vskip 0.5 cm 
$^{\rm b}$
Universit\'e de Lyon, F-69622, Lyon, France;
Universit\'e Lyon 1, Villeurbanne;\\
CNRS/IN2P3, UMR5822, Institut de Physique Nucl\'eaire de Lyon
\end{center}

\vskip 1.0 cm 
\begin{center} {\bf Abstract} \end{center}
  A particle-hole model is developed to describe the excitation
  spectrum of single \La hypernuclei and the possible presence of
  collective effects is explored by making a comparison with the
  mean-field calculations.  Results for the spectra of \carla, \oxyla,
  \cala, \zrla and \pbla hypernuclei are shown.  The
  comparison with the available experimental data is satisfactory. We
  find that collective phenomena are much less important in
  hypernuclei than in ordinary nuclei.

\vskip 1.5 cm 
%\newpage
\section{Introduction}
\label{sec:intro}

The properties of hypernuclei have been widely studied in several
experiments and today they are the object of a rich experimental
program.  The first experiments were aimed at identifying the
hypernuclei and at determining the energy of the hyperon embedded in
the nucleus. Presently, the improvement of the experimental techniques
allows the measurement of high-quality excitation spectra, which have 
been studied in ($\pi^+$,$K^+$) \cite{pil91,has96,aji98,hot01a},
($K^-$,$\pi^-$) \cite{tam94,agn05,agn05b} or ($e,e'K^+$)
\cite{miy03,gar05,iod07} reactions.  In the future, further advance in
these studies will be gained by means of $\gamma$-ray spectroscopy,
which will become a standard investigation tool \cite{has06,ma07}.

From the theoretical point of view, single-\La hypernuclei have been
studied within the mean-field (MF) approximation: the hypernucleus is
seen as a many-body system containing an impurity, and its properties
are solely determined by those of the hyperon moving in an average
potential generated by the interaction with the nucleons.  When
applied to various medium heavy nuclei, the MF model is quite
successful in reproducing the single hyperon energies, which are
single particle properties of the system.  On the other hand, the
excitation spectra of hypernuclei are, in general, related to the
system as a whole.

In purely nucleonic nuclei there are large differences between the
experimental excitation spectrum and that predicted by the MF model,
where the excitation energies are simply given by the differences
between the energies of the particle and hole states.  This indicates
that the many body-terms of the hamiltonian neglected in the MF
approximation play a relevant role.  It is therefore interesting to
study also the excitation spectra of hypernuclei by using models
beyond the simple MF approach.

In the past this kind of studies have been done within the Tamm
Dancoff Approximation (TDA) \cite{aue77,lop95}, which describes the
many-body excited state as a linear combination of particle-hole
excitations of a MF ground state. The theoretical inconsistencies of
the TDA theory are overcome by the Random Phase Approximation (RPA)
theory which considers some ground state correlations \cite{row70}.

We have developed an RPA-like model which describes the spectra of
single \La hypernuclei. In our model, the hypernucleus is formed when
a nucleon, from its state below the Fermi surface, a hole state, is
transformed into a \La which lies in one specific particle
state. The excited state of the hypernucleus is
a combination of various particle-hole excitations. This combination
is ruled by the residual \La-nucleon interaction.  The relevance of
the residual interaction, and therefore the possible presence of
collective phenomena, is investigated by comparing MF results with
those obtained by our model.

In the next section we present the details of the model. In Sect.
\ref{sec:calculations} we describe how we have chosen the input of our
calculations, namely the set of single particle wave functions and
energies, and the effective \La-nucleon interactions.  A selected set
of results concerning hypernuclei obtained from doubly-closed shell
nuclei is discussed in Sect.  \ref{sec:results}. Finally, in Sect.
\ref{sec:conclusions} we summarize the basic findings of our study and
draw our conclusions.

\section{The formalism}
\label{sec:formalism}

The starting point of our model is the MF description of the many-body
system. Each baryon is described by its single particle wave function
$\phi_\alpha$ and energy $\epsilon_\alpha$, where $\alpha$ indicates
the set of quantum numbers which identify the single particle level.

The definition of the single particle basis allows us to write the
many-body hamiltonian as \cite{row70}
\beq
H=\sum_{\alpha} 
   \epsilon_{\alpha} a^\dag_{\alpha} a_{\alpha}
- \frac 1 2 \sum_{h h'} \vbar_{h h' h h'} 
+ \frac 1 4 \sum_{\alpha \alpha' \beta \beta'}
   \vbar_{\alpha \alpha'\beta \beta'  } 
   N \left[ a^\dag_{\alpha} a^\dag_{\alpha'} a_{\beta'}  a_{\beta}
     \right]
\,\,\,,
\label{eq:hamiltonian}
\eeq
where we have indicated with $a^\dag$ and $a$ the creation and
annihilation operators, and with $N$ the normal ordered product
operator \cite{fet71}. In the above equation, the bar over the
interaction $V$ indicates that both direct and exchange terms are
considered, the $h$ label identifies a state below the Fermi surface,
while the Greek subindexes refer to states above or below the Fermi
surface.

We describe the formation of the hypernucleus as a time dependent
fluctuation of the MF ground state. We solve the time dependent
Schr\"odinger equation by using the variational principle
\beq
\delta 
\left\{
<\Psi(t)| \left[ H - i \hbar \frac {\partial}{\partial t} 
\right] |\Psi(t)>
\right\} = 0
\,\,.
\label{eq:var1}
\eeq

We search for the minimum of the above energy functional 
within the Hilbert subspace spanned by trial wave functions of the
form
\beq
|\Psi(t)> \rightarrow  
|\Phi(t)> =  e^{-i  E_0  t/ \hbar} 
\exp \left[\sum_{\Lambda h} C_{\Lambda h}(t) 
a^\dag_{\Lambda}a_h \right]
|\Phi_0>
\,\,,
\label{eq:psit}
\eeq
where $|\Phi_0>$ indicates the MF ground state of the system, formed
by nucleons only.  In equation (\ref{eq:psit}) we have indicated with
$a^\dag_{\Lambda}$ the operator which creates a \La, and with $a_h$
the operator annihilating a nucleon.  The coefficients $C_{\Lambda
  h}(t)$ are  complex numbers.

By using the ansatz (\ref{eq:psit}) in Eq. (\ref{eq:var1})
we express the variational equation as
\beq
\frac{\partial}{\partial C^*_{\Lambda h}(t)  }
<\Phi(t)| 
\left[ H - i \hbar \frac {\partial}{\partial t}  \right]
|\Phi(t)>
= 0
\label{eq:var2}
\,\,.
\eeq

By making a power expansion of the exponential of Eq. (\ref{eq:psit}),
and retaining only the terms up to the second order in $C(t)$, we
obtain the expression \cite{row70}
\begin{eqnarray}
\nonumber
 C_{\Lambda h}(t) (\epsilon_\Lambda - \epsilon_h) &~&
 +  \sum_{\Lambda' h'}   C^*_{\Lambda' h'}(t)
    \vbar_{\Lambda \Lambda' h h'} 
 +  \sum_{\Lambda' h'}   C_{\Lambda' h'}(t)
    \vbar_{\Lambda h' h \Lambda'} \\
&~&= i \hbar \frac{d}{dt}  C_{\Lambda h}(t)
\,\,.
\end{eqnarray}

Since we are looking for oscillating solutions, we write
the coefficients as
\beq
 C_{\Lambda h}(t) = W_{\Lambda h} e^{-i \nu t}
                  + Z^*_{\Lambda h} e^{ i \nu t}
\,\,,
\label{eq:freq}
\eeq
and by equating positive and negative frequencies components we
obtain the equations
\begin{eqnarray}
 W_{\Lambda h}  (\epsilon_\Lambda - \epsilon_h)
 +  \sum_{\Lambda' h'} \vbar_{\Lambda h' h \Lambda'} W_{\Lambda' h'}
 +  \sum_{\Lambda' h'} \vbar_{\Lambda \Lambda' h h'} Z_{\Lambda' h'}
= \hbar \nu W_{\Lambda' h'} 
\label{eq:w1}
\\
 Z^*_{\Lambda h}  (\epsilon_\Lambda - \epsilon_h)
 +  \sum_{\Lambda' h'} \vbar_{\Lambda h' h \Lambda'} Z^*_{\Lambda' h'}
 +  \sum_{\Lambda' h'} \vbar_{\Lambda \Lambda' h h'} W^*_{\Lambda' h'}
= - \hbar \nu Z^*_{\Lambda' h'} 
\,\,.
\label{eq:z1}
\end{eqnarray}

Introducing the matrix elements
\begin{eqnarray}
 H_{\Lambda h \Lambda'h'}  &=&
(\epsilon_\Lambda - \epsilon_h) \delta_{\Lambda \Lambda'} \delta_{h
  h'} + \vbar_{\Lambda h' h \Lambda'} 
\label{eq:vrpa7}
\\
 I_{\Lambda h \Lambda'h'} &=&
\vbar_{\Lambda \Lambda' h h'} 
\label{eq:vrpa8}
\,\,,
\end{eqnarray}
equations (\ref{eq:w1}) and (\ref{eq:z1}) 
can be expressed in the following matrix form 
\beq
\left(
   \begin{array} {cc}
   H  & I \\
  -I^\star & -H^\star \\ 
   \end{array} 
   \right)
  \left( 
  \begin{array}{c} W \\ Z \\
   \end{array}
  \right)
= \omega 
  \left( 
  \begin{array}{c} W \\ Z \\
   \end{array}
  \right)
\,\,,
\label{eq:rpasec}
\eeq
where we have defined $\omega = \hbar \nu$.  We have rewritten the
secular equations in a matrix form to show their similarity with the
traditional RPA secular equations \cite{rin80}. Our derivation of the
secular equations shows that the $W$ and $Z$ amplitudes are introduced
on equal footing, and that the choice of neglecting one of the two
terms of Eq. (\ref{eq:freq}) is a further approximation of the
theory. On the other hand, the derivation of the secular equations
done with the equation of motion method \cite{row70,rin80} indicates
that the $Z$ amplitudes are related to ground state correlations.
Neglecting the $Z$ amplitudes produces TDA secular equations which
have been used in the past \cite{aue77,lop95} to describe hypernuclei.
A discussion on the relevance of the ground state correlations
introduced by the $Z$ amplitudes will be done at the beginning of
section \ref{sec:results}. However, we would like to point out here
that the ground state correlations we include are only those described
by \La-nucleon excitations, and not by nucleon-nucleon
excitations. For this reason our model does not include nucleonic
correlations as it is done in the traditional RPA theory.

In our calculations the ground state of the nucleus and the final
state of the hypernucleus have definite angular momentum values.  For
this reason we rewrite the above equations in angular momentum
coupling scheme. The angular momentum coupled amplitudes are defined
in terms of the uncoupled ones as
\begin{eqnarray}
W^J_{\Lambda h}&=&\sum_{m_\Lambda m_h}<j_\Lambda m_\Lambda j_hm_h|JM>
W_{\Lambda h} \\
Z^J_{\Lambda h}&=&\sum_{m_\Lambda m_h}
<j_\Lambda m_\Lambda j_hm_h|J-M>(-1)^{J-M} 
Z_{\Lambda h} 
\,\,,
\end{eqnarray}
where the $j$'s indicate the total angular momentum 
characterizing the single particle wave functions and the $m$'s their
z-axis components. 

We express the terms (\ref{eq:vrpa7} - \ref{eq:vrpa8}) of the secular
matrix (\ref{eq:rpasec}) as
\begin{eqnarray}
H^J_{\Lambda h \Lambda'h'} &=& 
(\epsilon_\Lambda - \epsilon_h) \delta_{\Lambda \Lambda'} \delta_{h
  h'} +
v^J_{\Lambda h \Lambda' h'} \\
I^J_{\Lambda h \Lambda'h'} &=& u^J_{\Lambda h \Lambda' h'}
\label{eq:ematend}
\,\,,
\end{eqnarray}
where we have defined
\begin{eqnarray}
&~& v^J_{\Lambda h\Lambda'h'} = 
v_{\Lambda h\Lambda'h'}^{J dir}-v_{\Lambda h\Lambda'h'}^{J exch}
\nonumber\\
&=& \sum_K (-1)^{j_h+j_{\Lambda'}+K} \hat{K}
\sixj {j_\Lambda} {j_h} {J} {j_{\Lambda'}}   {j_{h'}} {  K}
\nonumber\\
&~& \left[
<j_\Lambda j_{h'}K||V||j_{h}j_{\Lambda'}K>-(-1)^{j_h+j_{\Lambda'}-K}
<j_{\Lambda}j_{h'}K||V||j_{\Lambda'}j_{h}K>
\right]
\,\,,
\label{eq:vj}
\end{eqnarray}
and
\begin{equation}
u^J_{\Lambda h\Lambda'h'}=
u_{\Lambda h\Lambda'h'}^{J dir}-u_{\Lambda h\Lambda'h'}^{J exch}
=(-1)^{j_{h'}-j_{\Lambda'}-J} v^J_{\Lambda h h'\Lambda'}
\label{eq:uj}
\,\,.
\end{equation}
Here we use the notation 
\beq
\hat{J}=\sqrt{2J+1}
\,\,,
\eeq
and the term in curly brackets is the Wigner $6j$ coefficient.
The double bar in the matrix elements in Eq. (\ref{eq:vj}) indicates 
that in the calculation of the angular part we consider the reduced
matrix element as defined by the Wigner-Eckart theorem \cite{edm57}.

In calculating the interaction matrix elements we use 
the Fourier transform of the interaction expressed in momentum 
space
\begin{eqnarray}
\nonumber
v(|\br_1-\br_2|) &=&  
\frac {1}{(2\pi)^3} \int d\bqu \, e^{i\bqu \cdot |\br_1-\br_2|} \\
&~& \left[   
       F_\Lambda(q) 
     + G_\Lambda(q) \,\bsigma(1) \cdot \bsigma(2) +
       H_\Lambda(q) \,  S_{12}(\bqu)  \right]
\,\,\,,
\label{eq:fourier}
\end{eqnarray}
where we have used $q\equiv|\bqu|$ and we have indicated with 
$\bsigma$ the Pauli spin matrices and with $S_{12}$ the 
usual tensor operator \cite{edm57}.
The above expression allows us to
separate the terms dependent on $\br_1$ from those dependent on
$\br_2$. The explicit expressions of the matrix elements for the
various terms of the interaction are given in the Appendix.
The solution of Eq. (\ref{eq:rpasec}) provides the excitation energies
$\omega$ and also the wave function of the excited state.  

We then consider the transition from the nucleus ground state to an
excited state of the hypernucleus induced by a generic one-body
operator $T_J$. The transition matrix element is given by
\begin{eqnarray}
\nonumber
<J|T_J|0> = 
\sum_{\Lambda h} &~&  
\Big[ W^J_{\Lambda h} <j_\Lambda||T_J||j_h> \\ 
&+&  
(-1)^{J+j_\Lambda -j_h}
Z^J_{\Lambda h} <j_h||T_J||j_\Lambda> \Big] 
\label{eq:transj1}
\,\,.
\end{eqnarray}

For the natural parity states we use the operator 
\beq
T_J = j_J(qr) Y_{JM}(\Omega)
\label{eq:tmenat}
\,\,,
\eeq
where $j_J(qr)$ are the spherical Bessel function,
and $Y_{JM}$ the spherical harmonics.
The matrix elements to be inserted in Eq. (\ref{eq:transj1})
are 
\beq
<j_a||T_J||j_b> = 
\int dr r^2 R_a^\ast(r) R_{b}(r) j_J(qr)
<j_a||Y_J||j_b> 
\,\,,
\eeq
where $R(r)$ is the radial part of the single
particle wave function and 
\begin{eqnarray}
\nonumber
<j_a||Y_J||j_b> &\equiv& 
<l_a \half j_a \| Y_J \| l_b \half j_b> \\
&=&
(-1)^{j_a+\half} \, 
\frac{\hat {j_a} \hat {j_b} \hat {J}} {\sqrt{4\pi}}
\threej {j_a}{J}{j_b}{\half}{0}{-\half} \xi(l_a+l_b+J)  
\,\,.
\end{eqnarray}
In the above equation we have used the Wigner 3-j symbol \cite{edm57},
and $\xi(l)=1$ if $l$ is even, and 0 otherwise.

For the unnatural parity states we use the operators 
\beq
T_{L=J\pm1} = j_L(qr)
\left[ Y_L(\Omega) \otimes \bsigma(1) \right]^J_M
\label{eq:tmeunat}
\,\,,
\eeq
therefore the matrix elements to be inserted in 
Eq. (\ref{eq:transj1}) are 
\begin{eqnarray}
\nonumber
&~&
<j_a||T_L||j_b> = \\
&~&
\int dr r^2 R_a^\ast(r) R_{b}(r) j_L(qr)
<j_a||\left[ Y_L(\Omega) \otimes \bsigma(1) \right]^J||j_b> 
\,\,,
\end{eqnarray}
with 
\begin{eqnarray}
\nonumber
<j_a \|[Y_{J+s} \otimes \bsigma ]^J \| j_b> &=& 
(-1)^{l_a+l_b+j_b+\half} \,
\frac{ \hat{j}_a \hat{j}_b} {\sqrt{4\pi}} \,
\frac{\chi_a + \chi_b + sJ + \delta_{s,1}}
                     {\sqrt{J+\delta_{s,1}}} \\
& &
\threej{j_a}{j_b}{J}{\half}{-\half}{0} \xi(l_a + l_b + J +1)
\,\,\,,
\end{eqnarray}
where we have defined 
\[
\chi = (-1)^{l+j+\half} \left( j+ \half \right)
\,\,,
\]
and $s=\pm1$.

We also consider the transition between two excited states of the
hypernucleus induced by a generic one-body operator $(EM)$. By using 
standard techniques \cite{rin80,suh07} we obtain for 
the transition matrix elements
\begin{eqnarray}
\nonumber 
&~&< \Phi_\nu|(EM)| \Phi_\nu' >  \\
\nonumber
&=&   
\sum_{\Lambda h} \sum_{\Lambda' h'} \delta_{hh'}
\left[ W^\nu_{\Lambda h} W^{\nu'}_{\Lambda'h'} 
       <\Lambda|(EM)|\Lambda'> -  
      Z^\nu_{\Lambda h} Z^{\nu'}_{\Lambda'h'} 
        <\Lambda'|(EM)|\Lambda> \right] \\
&-&  
\sum_{\Lambda h} \sum_{\Lambda' h'} \delta_{\Lambda\Lambda'}
\left[ W^\nu_{\Lambda h} W^{\nu'}_{\Lambda' h'} <h'|(EM)|h> -  
       Z^\nu_{\Lambda h} Z^{\nu'}_{\Lambda' h'} <h|(EM)|h'> \right] 
\label{eq:tgamma}
\,\,.
\end{eqnarray}

We calculate the electromagnetic transition probabilities 
between two states of the
hypernucleus by using the expression \cite{bla52} 
\beq
{\cal T}^L_{if}= \frac {4 k}{\hbar c^2 (2J_i+1) (2L+1)} 
\left| \langle J_f || (EM)_L || J_i \rangle  \right|^2
\eeq
where the indexes $i$ and $f$ indicate, respectively, the initial and
final state of the transition, $k$ is the modulus of the emitted
photon momentum, and $(EM)_L$ is the electromagnetic operator, of
multipolarity $L$.  In the above equation, as before, the double bar
indicates that we have to evaluate the reduced matrix element of the
angular dependent part.

We find that, in the angular momentum coupling scheme, the reduced
transition matrix element between the two excited states {\sl i.e.} Eq.
(\ref{eq:tgamma}), can be written as 
\begin{eqnarray}
&~& <J||(EM)_L||J'> = \hat{J} \hat{J'}
\sum_{\Lambda h} \sum_{\Lambda' h'} \Bigg\{ \delta_{hh'}
\sixj {J'}{L}{J}{j_\Lambda}{j_h}{j_{\Lambda'}} 
\nonumber \\
&~&
\Big[
 (-1)^{j_\Lambda+j_h+J'+L} 
    W^J_{\Lambda h} W^{J'}_{\Lambda'h'} <j_\Lambda||(EM)_L||j_{\Lambda'}>
\nonumber \\
&~&
-(-1)^{j_{\Lambda'}+j_{h'}+J'} 
     Z^J_{\Lambda h} Z^{J'}_{\Lambda'h'} <j_{\Lambda'}||(EM)_L||j_\Lambda>
\Big]
%\Bigg\}
\nonumber \\
&~& 
%\sum_{\Lambda h} \sum_{\Lambda'h'} \Bigg\{ 
-\delta_{\Lambda\Lambda'} \sixj {L}{J'}{J}{j_\Lambda}{j_h}{j_{h'}} 
\nonumber \\
&~&
\Big[
 (-1)^{j_{\Lambda'}+j_{h}+J} W^J_{\Lambda h} W^{J'}_{\Lambda'h'} <j_{h'}||(EM)_L||j_h>
\nonumber \\
&~&
-(-1)^{j_{\Lambda}+j_{h'}+J'+L} Z^J_{\Lambda h} Z^{J'}_{\Lambda'h'} <j_{h}||(EM)_L||j_{h'}>
\Big]
\Bigg\}
\label{eq:tgammaj}
\,\,\,.
\end{eqnarray}

The expressions of the single particle matrix elements for natural
parity (electric), and unnatural parity (magnetic) transitions, are
given in Ref. \cite{boh69}.

\section{Details of the calculations}
\label{sec:calculations}

We have applied the formalism described in the previous section to
calculate the excitation spectrum of single \La hypernuclei formed by
the excitation of nucleonic doubly closed shell nuclei. 
Specifically, we have investigated the \carla, \oxyla, \cala, \zrla and
\pbla hypernuclei. 

%----------------------------------------------------------------------
% gaussian parameters
%----------------------------------------------------------------------
\begin{table}[ht]
\begin{center}
\begin{tabular}{cccrrr}
\hline
       &        & $c$ & $\sigma$ & $t$ \\
\hline
 LN$\delta$ & & 20.0  & 70.0 & -28.0 & \\
\hline
       & $w_1$ & -4.5   & -1.26  &  15.0 \\
 LND & $w_2$ & -104.0 & -7.56  & 145.0 \\
       & $w_3$ & 590.0  &  182.0 &   1.0 \\
\hline
        & $w_1$ &   0.0 & -1.0 &   8.0 \\
 LNDE & $w_2$ & -92.0 & -3.5 &  50.0 \\
        & $w_3$ & 518.0 & 90.0 & 100.0 \\
\hline
        & $\beta_1$ & 1.5  & 1.5 & 1.0 \\
        & $\beta_2$ & 0.9  & 0.9 & 0.6 \\
        & $\beta_3$ & 0.5  & 0.5 & 0.4 \\
\hline
\end{tabular}
\end{center}
%\bigskip
\caption
{\small
Values of the parameters of the three \La-nucleon interactions
 used in this work. 
The LN$\delta$ interaction is of zero range
type, and, in the table, we indicate the values of the $U^{c,\sigma,t}$
parameters of  Eq. (\ref{eq:forcezero}) in MeV fm$^3$.
The other two interactions have been constructed as indicated by 
Eqs. (\ref{eq:gauss1}) and (\ref{eq:gauss2}). The values of the $w$
coefficients for the scalar ($c$) and spin ($\sigma$) channels are
expressed in MeV, those of the tensor channel ($t$) in MeV
fm$^{-2}$. The ranges of the gaussians $\beta_i$ are the same for both 
LND and LNDE interactions and their values are expressed in fm.
}
\label{tab:3r_gauss}
\end{table}

%----------------------------------------------------------------------
% 16O_Lambda spectrum
%----------------------------------------------------------------------
\begin{table}[ht]
\begin{center}
\begin{tabular}{c|c|ccccc}
\hline
\multicolumn{7} {c} {\oxyla} \\
\hline
 $J^\pi$ & p h   & MF & LN$\delta$ & LND  & LNDE & exp \\
\hline
  $0^-$ & $(1s_{1/2})_\Lambda (1p_{1/2})^{-1}_n$  
          & 0.0  &   0.0  & 0.0    & 0.0    & 0.0  \cite{uka07} \\
  $1^-_1$ & $(1s_{1/2})_\Lambda (1p_{1/2})^{-1}_n$  
          & 0.0  &  0.71  & 0.71   & 0.69  & 0.026  \cite{uka04} \\
  $1^-_2$ & $(1s_{1/2})_\Lambda (1p_{3/2})^{-1}_n$  
          & 6.16 &  6.58  & 6.55   & 6.56  & 6.562  \cite{uka07}\\
  $2^-$ & $(1s_{1/2})_\Lambda (1p_{3/2})^{-1}_n$  
          & 6.16 &  6.77  & 6.76   & 6.79  & 6.784  \cite{uka07} \\
  $1^+$ &  $(1p_{3/2})_\Lambda (1p_{1/2})^{-1}_n$      
          & 9.97 & 10.25 & 10.10  & 10.36  &  \\
  $2^+$ &  $(1p_{3/2})_\Lambda (1p_{1/2})^{-1}_n$    
          & 9.97 & 10.33 & 10.39  & 10.43  &  10.57 \cite{has98}\\
  $0^+$ &  $(1p_{1/2})_\Lambda (1p_{1/2})^{-1}_n$   
          &11.17 & 11.10 & 10.40  & 10.55  &  \\
\hline
\end{tabular}
\end{center}
%\bigskip
\caption
{\small 
Energies in MeV of the \oxyla spectrum, obtained in MF
approximation, and with the three interactions used in our work,
compared to the experimental values given in
Refs. \cite{uka04,uka07,has98}. The energy resolution obtained in
\cite{uka04,uka07} is of few keV, while that of the 2$^+$ state is of
2 MeV \cite{has98}. The MF particle-hole components are also shown.
}
\label{tab:16ospectrum}
\end{table}

In analogy to what is done in the application of the Landau-Migdal
theory \cite{mig67} to finite nuclear systems \cite{spe77,gru06}, the
mean field is described by a phenomenological potential, in our case a
spherical Woods-Saxon well, of the form  
\begin{eqnarray}
\nonumber
U^t(r) &=&
\frac{-V^t_0}
{1+ \exp { \left[ \Big( r-R^t_0 \Big) / a^t_0 \right] }} \\
&-&
\left[\frac{\hbar c}{m_{\pi}c^2}\right]^2 \,
\left( \frac {V^t_{ls}}{a^t_{ls}\,r} \right) \,
\frac{\exp{\left[ \Big( r-R^t_{ls} \Big) / a^t_{ls} \right] }}
{ \left\{ 
1+ \exp \left[ \Big( r-R^t_{ls} \Big) / a^t_{ls} 
      \right] \right\}^2 } 
\,\, {\bf l}\cdot \bsigma \, - V^t_{C}(r)
\,\,,
\label{eq:fin-wspot}
\end{eqnarray} 
where $m_{\pi}$ is the pion mass and the Coulomb term $V^t_C(r)$,
active only for protons, is that produced by a
homogeneous charge distribution. 

%%%%%%%%%%%%%%%%%%%%%%%%%%%%%%%%%%%%%%%%%%%%%%%%%%%%
% Figure Force
%%%%%%%%%%%%%%%%%%%%%%%%%%%%%%%%%%%%%%%%%%%%%%%%%%%%
\begin{figure}[ht]
\begin{center}
\includegraphics[scale=0.4, angle=0] {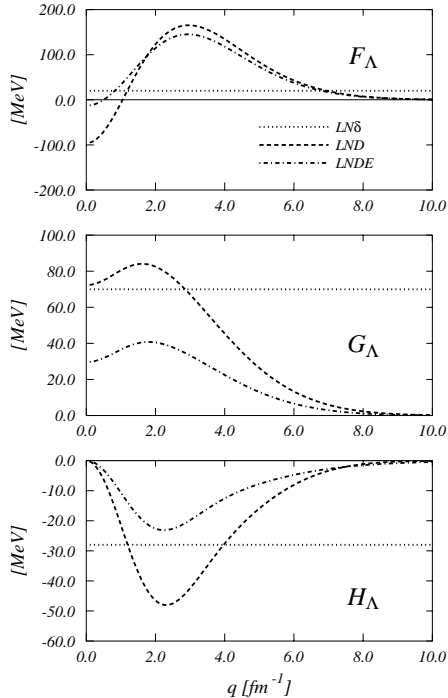}
\vskip -0.5 cm  
\caption{\small Momentum dependence of the 
\La - nucleon interactions for each channel of Eq. (\ref{eq:fourier}).
The $LN\delta$ zero-range interaction is constant in momentum space.
The parameters of the other two interactions are given in Tab. 
\ref{tab:3r_gauss}.
}
\label{fig:force}
\end{center}
\end{figure}

For the nucleonic part we used parameterizations chosen to reproduce
the single particle energies of the levels close to the Fermi surface
and the r.m.s. charge radii. The parameters of these potentials have
been taken from the literature \cite{ari07} for the \car, \oxy, \ca
and \pb nuclei. For the \zr nucleus we used $V^p_0$=55.88 MeV,
$V^n_0$=48.12 MeV, $V^p_{ls}=V^n_{ls}$=7.7 MeV, and
$a^{p,n}_{0}=a^{p,n}_{ls}$=0.645 fm, and $R^{p,n}_0$=5.695 fm,
$R^{p,n}_{ls}$=5.68 fm.

In our calculations also the \La mean field potential is described by
a spherical Woods-Saxon well with a set of parameters values very
close to those given in \cite{mil88}.  We used $V^\Lambda_0$= 29.0
MeV, $V^\Lambda_{ls}$=2.0 MeV, and $a^\Lambda_0= a^\Lambda_{ls}$=0.6
fm for all the hypernuclei, and we changed the radii in order to
reproduce the ground state binding energies of the various
hypernuclei. For \carla, \oxyla, \cala, \zrla and \pbla we used
$R^\Lambda_0=R^t_{ls}$=2.50, 2.73, 4.00, 5.60, 7.60 fm, respectively.

In both the nucleonic and \La sectors the mean-field potential is
diagonalized in a truncated harmonic oscillator basis. This procedure
discretizes automatically the continuum.  Since we are interested in
the lowest excited states a proper description of the escape width is
not necessary \cite{spe77}.  In Sect.  \ref{sec:results} we discuss
results obtained for excited states dominated by particle-hole
transitions where the \La is in a bound state.

Our calculations have been done with a configuration space which
considers 54 single particle levels describing the \La hyperon, up to
the $5p_{1/2}$ level. This corresponds to consider 10 major harmonic
oscillator shells.  A comparison with results obtained with a smaller
configuration space of only 22 states, up to the $3p_{1/2}$ level,
shows differences in the eigenvalues of about 10, 20 keV.

As already indicated in Eq. (\ref{eq:fourier}), the residual
interaction between \La and the nucleons is considered to be a
sum of scalar, spin and tensor terms 
\begin{eqnarray}
\nonumber
V_{\Lambda N} = V^c_{\Lambda N}(r) 
 &+& V^\sigma_{\Lambda N}(r) \, \bsigma(\Lambda) \cdot \bsigma(N) \\
 &+& V^t_{\Lambda N}(r) \left[ 3 
   \frac  {(\bsigma(\Lambda) \cdot \br)( \bsigma(N)\cdot \br)}{\br^2} 
   - \bsigma(\Lambda) \cdot \bsigma(N) \right]
\label{eq:force}
\,\,\,.
\end{eqnarray}
In the above equation $N$ indicates the nucleon and 
the $V^{c,\sigma,t}_{\Lambda N}(r)$ terms are the
Fourier transforms of the $F_\Lambda$, $G_\Lambda$ and $H_\Lambda$
terms of Eq. (\ref{eq:fourier}). With respect to the \La-nucleon
interactions commonly used in shell-model calculations \cite{has06},
we neglect the spin-orbit term. The contribution of this term is
rather small \cite{mil05}, and we consider it at the mean-field level
by including a spin-orbit term in the Woods-Saxon potentials. The
effects of the spin-orbit terms of the residual interaction have been
studied in self-consistent RPA calculations for ordinary nuclei
\cite{sil06} and they have been found to be rather small.

The sensitivity of our results to the \La-nucleon interaction has been 
tested by comparing the results obtained with three different
parameterizations of the $V^{c,\sigma,t}_{\Lambda N}(r)$ functions.
In a first parameterization, which we call LN$\delta$, the three
functions are assumed to be of zero-range type 
\beq
V^{c,\sigma,t}_{\Lambda N}(r) = U^{c,\sigma,t}_{\Lambda N} \delta(r)
\label{eq:forcezero}
\,\,.
\eeq
The values of the constants $U^{c,\sigma,t}$ are given in the first
row of Tab. \ref{tab:3r_gauss}.

For the other two interactions, in analogy to the YNG interaction
\cite{yam85,ban90}, we parametrize each radial term of the force
(\ref{eq:force}) as a sum of three gaussians, specifically,
\begin{equation}
V^{c,\sigma}_{\Lambda N}(r) = 
\sum_{i=1,3} w_i^{c,\sigma} \exp[{-(r/\beta_i^{\,c,\sigma})^2}]
\,\,\,,
\label{eq:gauss1}
\end{equation}
for the scalar and spin terms, and  
\begin{equation}
V^{t}_{\Lambda N}(r) = 
\sum_{i=1,3} w_i^t r^2 \exp[{-(r/\beta^t_i)^2}]
\,\,\,,
\label{eq:gauss2}
\end{equation}
for the tensor term.  

The interaction LN$\delta$ and the finite range interaction LND are
used in calculations where only the direct terms of the interaction
matrix elements, see Eq.  (\ref{eq:vj}), are considered, while we
consider both direct and exchange terms when we use the finite range
interaction we call LNDE.  The values of the parameters of each
interaction have been chosen to reproduce at best the energies of the
first 0$^-$, 2$^-$ and 1$^-$ excited states of \oxyla. These
parameters values are given in Tab. \ref{tab:3r_gauss} and the
momentum dependence of the three interactions is shown in Fig.
\ref{fig:force}.  In Tab.  \ref{tab:16ospectrum}, we compare the
results of our calculations in \oxyla with the experimental values
presented in Refs. \cite{has98,uka04,uka07}.  In the table all
the energies have been renormalized to the energy of the lowest state,
which, in our calculations, is always the 0$^-$.

We have also calculated the spectrum of \oxyla with the YNG
interaction of Refs. \cite{yam85,ban90}. This force reproduces the
G-matrix interaction built on a Nijmegen \La-nucleon potential
\cite{nag79}.  The spectrum we have obtained with this interaction, is
rather different from the experimental one. The lowest energy state is
a 1$^+$ followed by a 1$^-$ at 0.49 MeV, a 2$^-$ at 13.34 MeV. The
first 0$^-$ state appears at 17.94 MeV. We expected such a bad result,
since our theory requires interactions which effectively consider also
many-body effects not included in the G-matrix calculations
\cite{nak84}.  For this reason a straightforward comparison between
our effective interactions with more microscopic interactions is not
appropriate.

\section{Results}
\label{sec:results}
%%%%%%%%%%%%%%%%%%%%%%%%%%%%%%%%%%%%%%%%%%%%%%%%%%%%
% Figure TDA RPA
%%%%%%%%%%%%%%%%%%%%%%%%%%%%%%%%%%%%%%%%%%%%%%%%%%%%
\begin{figure}[ht]
\begin{center}
\includegraphics[scale=0.6, angle=90] {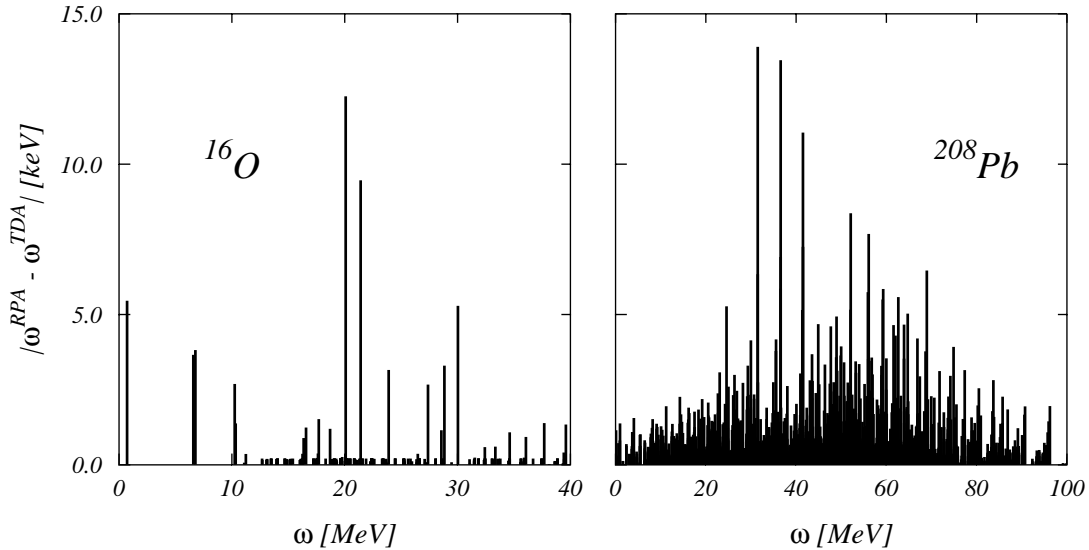} 
\caption{\small Differences between RPA-like and TDA energy eigenvalues 
of \oxyla and \pbla hypernuclei calculated with the LN$\delta$
interaction.}
\label{fig:rpatda}
\end{center}
\end{figure}

Before presenting the results obtained by applying our model to
\carla, \oxyla, \cala, \zrla and \pbla hypernuclei, we make a comment
about the presence of some ground state correlations in our model.  In
our approach, these correlations are taken into account by the $Z$
amplitudes. These effects are related to the RPA-like description of
the ground state \cite{row70}, and they describe a limited set of
correlations. For example they do not describe the short-range
correlations \cite{ari01}, and they do not consider the partial
occupation probability of single particle states both below and above
the Fermi surface. We have already pointed out that in our model 
nucleonic correlations are not considered.

%%%%%%%%%%%%%%%%%%%%%%%%%%%%%%%%%%%%%%%%%%%%%%%%%%%%
% Figure FF 16O
%%%%%%%%%%%%%%%%%%%%%%%%%%%%%%%%%%%%%%%%%%%%%%%%%%%%
\begin{figure}[ht]
\begin{center}
\includegraphics[scale=0.5, angle=0] {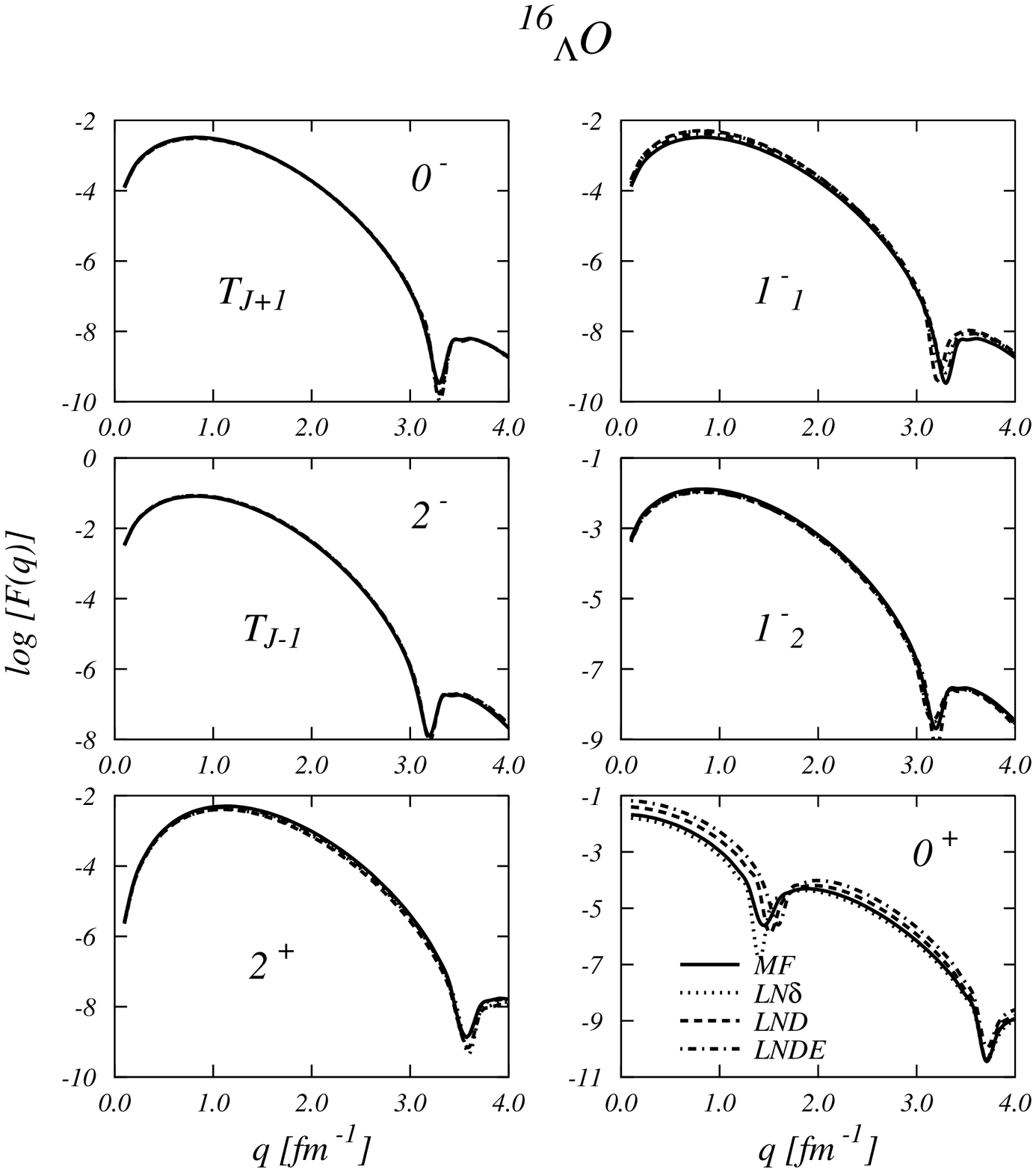} 
\vskip -0.3 cm 
\caption{\small Transition form factors, Eq. (\ref{eq:ffq}), obtained
  in MF approximation and calculated with the three 
  residual interactions used in this work, as a
  function of the momentum $q$ of the operators (\ref{eq:tmenat}) and
  (\ref{eq:tmeunat}).  For the unnatural parity excitations 0$^-$
  and 2$^-$, the operator inducing the transition is indicated,
  see Eq.  (\ref{eq:tmeunat}).  The subindexes 1 and 2 of the 1$^-$
  states indicate the lower and higher energies states respectively,
  see Tab. \ref{tab:16ospectrum}.  }
\label{fig:fmo16}
\end{center}
\end{figure}

A quantitative indication of the relevance of these RPA-like ground
state correlations, can be obtained by comparing our results with
those of the TDA theory obtained by solving Eq.  (\ref{eq:w1}) without
the term which multiplies the $Z$ amplitude.  Since we work with a
discrete configuration space, the number of eigenstates corresponding
to positive energies obtained in RPA and TDA is the same. For this
reason we can immediately identify the energy difference of each
eigenstate.  As an example of the results we have obtained, we show in
Fig.  \ref{fig:rpatda} these energy differences for the \oxyla and
\pbla hypernuclei as a function of the RPA energy.  These results have
been obtained by using the LN$\delta$ interaction.  In the left panel
of the figure we show 273 \oxyla energy differences corresponding to
all the solutions we found for all the positive and negative parity
multipole excitations up to $J=4$. In the case of \pbla we show 5024
energies differences corresponding to multipole excitations up to
$J=10$.

We observe that only in few cases the energy differences are larger
than 10 keV. Specifically, for the states of our interest, below 10
MeV for \oxyla and below 4 MeV for \pbla, these differences are even
smaller. The results of our RPA-like theory and those of the TDA are
quantitatively equivalent. However, since the RPA is the complete and
consistent theory describing linear combination of one-particle,
one-hole excitations \cite{row70}, we used our RPA-like model to
obtain all the results presented in the remaining part of the article.

We start the presentation of our results by showing in Tab.
\ref{tab:16ospectrum} the spectrum of \oxyla.  In this table also the
MF results and the experimental values of Refs.
\cite{has98,uka04,uka07} are shown.  The energy differences between
the 0$^-$, 2$^-$ and 1$^-_2$ states have been used to adjust the
parameters of the \La-nucleon forces. These parameterizations are not
unique.  We have found sets of parameters values which produce
very collective 0$^+$ states at low energy, in some case this was even
the lowest energy state. While there are no experimental evidences of
such low energy 0$^+$ state, there are indications of a 0$^+$ state
around 10-11 MeV \cite{bru78,dal86}. For this reason we have used
forces predicting a 0$^+$ state around 11 MeV. The energies of the
other states are a prediction of our model.

The MF energies are already rather close to the experimental values
and our RPA-like results change them by about 4\% on average. The
experimental splitting between the two 1$^-$ states is not well
reproduced. Better agreement could be obtained with a rearrangement of
the force parameters, or by including additional force terms such as
the spin-orbit or velocity dependent terms.

The structure of the 0$^+$ state is somehow more collective than those
of the other states of the spectrum.  This feature becomes more
evident by observing the transition matrix elements from the ground
state to the hypernuclear excited states. In Fig. \ref{fig:fmo16} we
show the transition form factors
\beq
F(q) = \left| < \Psi_0|T(q)| \Psi_\nu > \right|^2
\label{eq:ffq}
\,\,,
\eeq
for the negative parity states of the spectrum and for the 0$^+$ and
the 2$^+$ states.  The transition operators are defined in Eqs.
(\ref{eq:tmenat}) and (\ref{eq:tmeunat}) for natural and unnatural
parity states respectively. For unnatural parity excitations, we use 
transition operators characterized by $L=J \pm 1$. In the case of
the 0$^-$ state only the $T_{J+1}$ is active, while for the 2$^-$, both 
operators are active. In this last case, however, we found that the
$T_{J+1}$ form factor is zero in mean-field approximation, and four
orders of magnitudes smaller than the $T_{J-1}$ form factor when the
residual interaction is active. For this reason we neglect it in our
discussion.  In Fig.  \ref{fig:fmo16} the transition form factors for
the two 1$^-$ states are labeled with the subindexes 1 and 2 to
indicate the lower and higher energy states respectively, see Tab.
\ref{tab:16ospectrum}.  Finally, in the lower panels we show the
transition form factors for the 2$^+$ and 0$^+$ states.

%----------------------------------------------------------------------
% 16O_Lambda transitions
%----------------------------------------------------------------------
\begin{table}[ht]
\begin{center}
\begin{tabular}{l|rrrrr}
\hline
\multicolumn{5} {c} {\oxyla} \\
\hline
        & MF & LN$\delta$ & LND  & LNDE  & \cite{mil07} \\
\hline
 M1  ($1^-_1 \rightarrow 0^-$) 
                 & 0.0    & 0.081   & 0.082    & 0.083  & 0.176 \\
 M1  ($1^-_2 \rightarrow 0^-$) 
                 & 0.433  & 0.406   & 0.406    & 0.404  & 0.336 \\
 M1  ($1^-_2 \rightarrow 1^-_1$) 
                 & 0.217  & 0.248   & 0.248    & 0.250  & 0.129 \\
 M1  ($2^- \rightarrow 1^-_1$) 
                 & 0.650  & 0.680   & 0.680    & 0.683  & \\
 M1  ($2^- \rightarrow 1^-_2$) 
                 & 0.0    & 0.050   & 0.049    & 0.048  & \\
 E1  ($0^+ \rightarrow 1^-_1$) 
                 & 0.231  & 0.232   & 0.309    & 0.329  & \\
 E1  ($0^+ \rightarrow 1^-_2$) 
                 & 0.0    & 0.012   & 0.002    & 0.006  & \\
\hline
\end{tabular}
\end{center}
%\bigskip
\caption
{\small
Electromagnetic transitions between various states of \oxyla, in
Weisskopf units, compared to the shell model results of
Ref. \cite{mil07}.
}
\label{tab:16gamma}
\end{table}

The behavior of these form factors as a function of $q$ is rather
interesting. Only the 0$^+$ transition shows a maximum at values of
$q$ near zero. For this reason substitutional experiments such as
($K^-,\pi^-$) are better suited to investigate the 0$^+$ states than
experiments which produce hypernuclei with a rather large value of the
momentum transfer, such as ($e,e'K^+$) \cite{ban90}.

In Fig. \ref{fig:fmo16} the MF transition form factors are compared
with those obtained in our RPA-like calculations with the three
interactions. For the cases we have investigated, we have found that
the presence of the interaction does not modify sensitively the MF
results, but for the 0$^+$ state. In this last case, while the
LN$\delta$ result is similar to the MF one, the two finite-range
interactions produce rather different form factors. In the MF model
the particle-hole excitation generating the lowest 0$^+$ state is the
$(1p_{1/2})_\Lambda (1p_{1/2})^{-1}_n$ transition. This is also the
main transition ($W$=0.998) obtained in the calculation with the
LN$\delta$ interaction. The situation changes slightly when the LND
interaction is used, since a relevant contribution of the
$(1p_{3/2})_\Lambda (1p_{3/2})^{-1}_n$ transition ($W=0.121$) is
associated to the still dominant $(1p_{1/2})_\Lambda
(1p_{1/2})^{-1}_n$ transition ($W$=0.987).  The calculation with the
LNDE interaction produces a 0$^+$ state with a rather different
structure. The main transition is now the $(1p_{3/2})_\Lambda
(1p_{3/2})^{-1}_n$ ($W$=0.843) with a large contribution of the
$(1p_{1/2})_\Lambda (1p_{1/2})^{-1}_n$ component ($W$=0.366) and also
of the $(1s_{1/2})_\Lambda (1s_{1/2})^{-1}_n$ component ($W$=-0.327).

We obtain another indication of the relative collectivity of the 0$^+$
state by studying the electromagnetic transitions between the excited
stated of \oxyla.  The B-values of these transitions are shown in Tab.
\ref{tab:16gamma}.  The transitions forbidden in MF calculations have
relatively small values when the residual interaction is switched on.
All the M1 transitions are modified on the second significant figure
by the residual interaction. We use three digits to show the
difference between the results obtained with the various interactions.
The situation changes when the 0$^+$ state is present. The B-values of
the ($0^+ \rightarrow 1^-_1$) transition are quite different in the
various calculations. The LN$\delta$ result is similar to that
obtained in MF approximation, while the values obtained with the LND
and LNDE interactions are 38\% and 48\% larger than the MF value.
This result is again a consequence of the collective structure of the
0$^+$ state.

In Tab. \ref{tab:16gamma} we compare some of our results with those
obtained in shell model calculations \cite{mil07}.  The order of
magnitude of the B-values is the same in the two calculations, but the
differences are remarkable. Our results are roughly a factor two
smaller for the $1^-_1 \rightarrow 0^-$ transition and a factor two
larger for the $1^-_2 \rightarrow 1^-_1$ transition. The difference
between the results for the other transition is 17\%. It would be
interesting to compare with experimental values. Unfortunately, while
the experimental identification of the energies of the spectrum is
quite accurate \cite{tam05}, the measure of the B-values of the
transition requires higher statistics. The technological improvements
are very promising and these measurements will become feasible in the
near future.

%----------------------------------------------------------------------
% 12C_Lambda spectrum
%----------------------------------------------------------------------
\begin{table}[ht]
\begin{center}
\begin{tabular}{c|c|cccccc}
\hline
\multicolumn{8} {c} {\carla} \\
\hline
 $J^\pi$ & p h   & MF & LN$\delta$ & LND  & LNDE & E369 & FINUDA  \\
\hline
  $1^-_1$ & $(1s_{1/2})_\Lambda (1p_{3/2})^{-1}_n$  
          & 0.0  & 0.0   & 0.0    & 0.0    & 0.0 & 0.0\\
  $2^-$   & $(1s_{1/2})_\Lambda (1p_{3/2})^{-1}_n$  
          & 0.0  & 0.26  & 0.24   & 0.27  & -   & - \\
  $1^-_2$ & & - & - & - & - & 2.63  & 2.54  \\
  $1^-_3$ & & - & - & - & - & 6.09  & 5.04  \\
  $0^+$   & $(1p_{3/2})_\Lambda (1p_{3/2})^{-1}_n$  
          & 10.03 & 9.61 & 8.39 & 8.28 & 8.12  & 7.14  \\
 $2^+$ & $(1p_{3/2})_\Lambda (1p_{3/2})^{-1}_n$  
          & 10.03 & 9.38 & 9.33 & 9.79  &  -  & 9.34 \\
 $1^+$ & $(1p_{3/2})_\Lambda (1p_{3/2})^{-1}_n$  
          & 10.03 & 9.69 & 9.58 & 9.94  & 11.00 & 11.21   \\
 $3^+$ & $(1p_{3/2})_\Lambda (1p_{3/2})^{-1}_n$  
          & 10.03 & 9.87 & 9.80 & 9.95  &  -  & - \\
\hline
\end{tabular}
\end{center}
%\bigskip
\caption
{\small
Energies in MeV of the \carla spectrum calculated in MF approximation
and with the three interactions used in our work. 
The experimental values of E369 \cite{hot01a} are those 
reported in \cite{has06}. 
The values of the FINUDA experiment \cite{agn05} have been rescaled
with respect to the lowest energy value, and the angular momentum and
parity assignment is our guess. The uncertainty in the
energy resolution of the E369 experiment is 1.45 MeV, and that of the 
FINUDA experiment of 1.29 MeV. The MF particle-hole transitions
are also shown. 
}
\label{tab:12cspectrum}
\end{table}

In Tab. \ref{tab:12cspectrum} we compare the \carla spectrum with the
energies measured in the E369 \cite{hot01a} and FINUDA \cite{agn05}
experiments. The uncertainty in the energy resolution of the E369
experiment is about 1.5 MeV, and that of the FINUDA experiment is 1.29
MeV. The angular momentum and parity assignment to the energies
measured in the FINUDA experiment is our guess.
%%%%%%%%%%%%%%%%%%%%%%%%%%%%%%%%%%%%%%%%%%%%%%%%%%%%
% Figure FF 12C
%%%%%%%%%%%%%%%%%%%%%%%%%%%%%%%%%%%%%%%%%%%%%%%%%%%%
\begin{figure}[ht]
\begin{center}
\includegraphics[scale=0.5, angle=0] {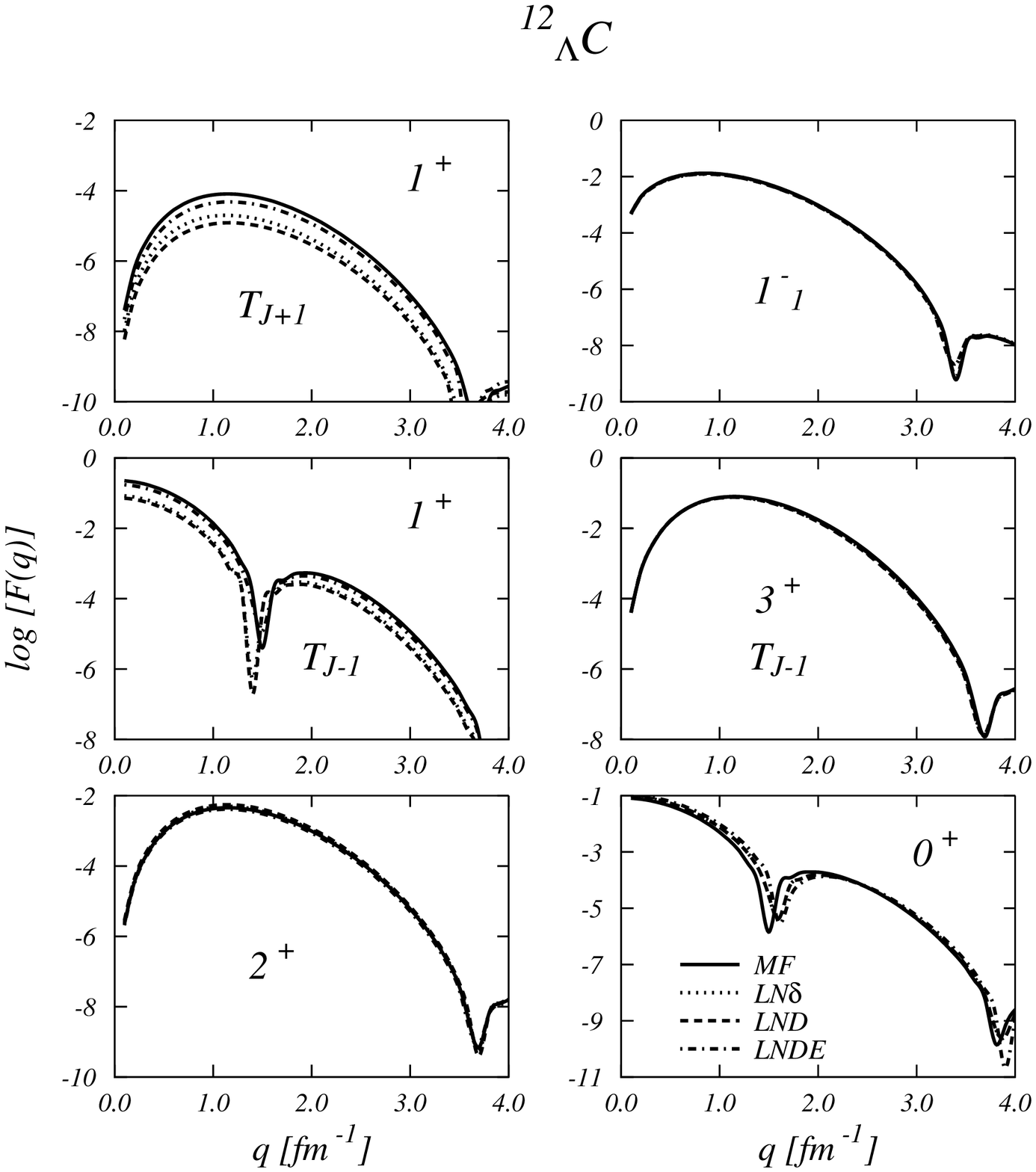} 
\vskip -0.5 cm 
\caption{\small Transition form factors as in Fig. \ref{fig:fmo16},
but for \carla. The energies of the various excited states are 
given in Tab. \ref{tab:12cspectrum}.  
}
\label{fig:fmc12}
\end{center}
\end{figure}

The presence in the experimental spectrum of two 1$^-$ states, that we
have called 1$^-_{2,3}$, is not predicted by our model. Shell model
calculations \cite{ito94,mil01} indicate that these states are
produced by neutron-\La transitions where the neutron is in the
$1p_{1/2}$ state, above the Fermi surface in the MF model.  The ground
state correlations of our approach do not consider partial occupation
probability of the single particle levels, therefore we are unable to
predict these excitations.  A description of \car as closed shell
nucleus is reasonable in kinematic regions where the detailed shell
structure is not relevant, such as giant resonances or the
quasi-elastic region, but for the case we are treating it is
inadequate. 
%----------------------------------------------------------------------
% 12C_Lambda transitions
%----------------------------------------------------------------------
\begin{table}[ht]
\begin{center}
\begin{tabular}{l|rrrrr}
\hline
\multicolumn{5} {c} {\carla} \\
\hline
        & MF & LN$\delta$ & LND  & LNDE  & \cite{dal78} \\
\hline
 M1  ($2^- \rightarrow 1^-_1$) 
                 & 0.0    & 0.082   & 0.082    & 0.082  & - \\
 E1  ($0^+ \rightarrow 1^-_1$) 
                 & 0.278  & 0.246   & 0.421    & 0.509  & 0.475 \\
 E1  ($2^+ \rightarrow 1^-_1$) 
                 & 0.139  & 0.135   & 0.121    & 0.158 & 0.133  \\
 E1  ($2^+ \rightarrow 2^-$) 
                 & 0.139  & 0.170   & 0.194    & 0.133 & 0.094  \\
\hline
\end{tabular}
\end{center}
%\bigskip
\caption
{\small
Electromagnetic transitions between various states of \carla, in
Weisskopf units, compared with 
the shell model results of Ref. \cite{dal78}.
}
\label{tab:12gamma}
\end{table}

Our model predicts the presence of excited states between 8 and 11
MeV. These states are all coming from the $(1p_{3/2})_\Lambda
(1p_{3/2})^{-1}_n$ excitation, and in MF approximation are degenerate
at 10.03 MeV. The residual interaction removes this degeneracy, and,
in general, the excitation energy of each state is lowered with
respect to the MF value. The 0$^+$ and 2$^+$ states are the lowest
ones. The LND and LNDE interactions give the 0$^+$ around 8.5 MeV,
where E369 finds a peak.  The identification of this experimental peak
with our 0$^+$ is however quite dubious, since the experiment produces
hypernuclei by using a probe that transfers high momentum values to
the system, while the 0$^+$ response has its maximum at small $q$
values. This is shown in the lowest right panel of Fig.
\ref{fig:fmc12}, where the transition form factor of this 0$^+$ state
is shown.

In our search for collective effects we have studied the transition
form factors of the various excitations, and, in Fig. \ref{fig:fmc12},
we show some of them. This figure is analogous to Fig.
\ref{fig:fmo16}, but for \carla. We have selected the transition form
factors of the positive states degenerate in MF, and the lowest 1$^-$
state. For the 3$^+$ state we show only the $T_{J-1}$ form factor,
since the $T_{J+1}$ is zero in the MF model.

The figure shows that the 2$^+$ and 3$^+$ transitions are essentially
unaffected by the effective interaction, as it is the ground state
1$^-$ transition. The 0$^+$ state, and, much more, the 1$^+$ state,
are instead rather sensitive to the presence of the residual
interaction.  The structure of the $1^+$ state is not collective. In
our calculation this state is described by the combination of two
particle-hole transitions, the $(1p_{3/2})_\Lambda (1p_{3/2})^{-1}_n$
transitions, which is the dominant one, and the $(1p_{1/2})_\Lambda
(1p_{3/2})^{-1}_n$ transition.  While in MF approximation this state
is described only by the first transition, the relative weight of the
two particle-hole components changes with the residual interaction.
The values of the $W$ amplitudes vary, respectively, from 0.99 and
0.10 for the LNDE interaction to 0.94 and 0.33 for LN$\delta$.  These
different couplings between the two states are sufficient to produce
the spreading of the form factors shown in Fig.  \ref{fig:fmc12}.

Also the RPA-like structure of the 0$^+$ state is mostly described by
the combination of the $(1p_{3/2})_\Lambda (1p_{3/2})^{-1}_n$ and
$(1s_{1/2})_\Lambda (1s_{1/2})^{-1}_n$ transitions. 
These
are the only two relevant transitions in the formation of the 0$^+$
state.  The relative weights of these two transitions change by
changing the residual interaction, and this produces the spreading of
the form factors shown in Fig. \ref{fig:fmc12}.

The B-values of some of the $\gamma$ transitions for the \carla
hypernucleus are shown in Tab. \ref{tab:12gamma}, and they are compared
with the shell model results of Ref. \cite{dal78}.  The $2^-
\rightarrow 1^-$ M1 transition is not present in the MF model, since
the two states are degenerate. Our RPA-like calculations remove this
degeneracy and produce B-values which are almost insensitive to the
different residual interactions.

The B-values of the E1 transitions $2^+ \rightarrow 1^-_1$ and $2^+
\rightarrow 2^-$ obtained in the MF model are identical, since the
wave functions of the initial and final states are the same, as it can
be deduced from the MF transitions shown in Tab.
\ref{tab:12cspectrum}.  The B-values obtained in our RPA-like
calculations are rather similar to the MF values. The situation
changes for the $0^+ \rightarrow 1^-_1$ E1 transition, where our
RPA-like results modify the MF value by 10, 70 and 100\% when the
LN$\delta$, LND and LNDE interactions are used, respectively.  As for
\oxyla, also in this case the structure of the $0^+$ state is
extremely sensitive to the residual interaction.  In our model this
state is described by a combination of the $(1p_{3/2})_\Lambda
(1p_{3/2})^{-1}_n$ transition, the dominant one, and the
$(1s_{1/2})_\Lambda (1s_{1/2})^{-1}_n$ transition. The coefficients of
these two particle-hole transitions are changed by the residual
interaction and this produces the changes in the B-values.  The
agreement between our results with those of the shell model
\cite{dal78} is more satisfactory than in the case of the \oxyla
hypernucleus.

In Tab. \ref{tab:40caspectrum} we show the results relative to the
spectrum of the \cala hypernucleus, for energies smaller than 10 MeV.
The changes of the RPA-like energies, with respect to those of the MF
calculations, are 3.5\%, on average. The only exception is the
$1_2^+$ state which shows changes around 20 \%.  We have analyzed the
transition form factors of this state, but we did not find any
remarkable effect which distinguishes it from the MF result.
%----------------------------------------------------------------------
% 40Ca_Lambda spectrum
%----------------------------------------------------------------------
\begin{table}[ht]
\begin{center}
\begin{tabular}{c|c|ccccc}
\hline
\multicolumn{7} {c} {\cala} \\
\hline
 $J^\pi$ & p h   & MF & LN$\delta$ & LND  & LNDE & \cite{tze02}  \\
\hline
  $1^+_1$ & $(1s_{1/2})_\Lambda (1d_{3/2})^{-1}_n$  
          & 0.0  & 0.0  & 0.0  & 0.0   & 0.22 \\
  $2^+_1$ & $(1s_{1/2})_\Lambda (1d_{3/2})^{-1}_n$  
          & 0.0  & 0.16 & 0.21 & 0.19  & 0.0 \\
  $0^+$ & $(1s_{1/2})_\Lambda (2s_{1/2})^{-1}_n$  
          & 2.60 & 2.50 & 2.74 & 2.42  & 2.92 \\
  $1^+_2$ & $(1s_{1/2})_\Lambda (2s_{1/2})^{-1}_n$  
          & 2.60 & 3.07 & 3.13 & 2.93  & 3.44 \\
  $2^+_2$ & $(1s_{1/2})_\Lambda (1d_{5/2})^{-1}_n$  
          & 6.00 & 6.04 & 6.08 & 6.09  & 6.19 \\
  $3^+$ & $(1s_{1/2})_\Lambda (1d_{5/2})^{-1}_n$  
          & 6.00 & 6.10 & 6.10 & 6.11  & 6.01 \\
  $2^-_1$ & $(1p_{3/2})_\Lambda (1d_{3/2})^{-1}_n$  
          & 7.81 & 7.72 & 7.69 & 7.77  & 7.98 \\
  $0^-$ & $(1p_{3/2})_\Lambda (1d_{3/2})^{-1}_n$  
          & 7.81 & 7.77 & 8.02 & 7.07  & 9.54 \\
  $1^-_1$ & $(1p_{3/2})_\Lambda (1d_{3/2})^{-1}_n$  
          & 7.81 & 7.95 & 7.91 & 8.05  & 8.78 \\
  $3^-$ & $(1p_{3/2})_\Lambda (1d_{3/2})^{-1}_n$  
          & 7.81 & 8.04 & 8.24 & 7.99  & 8.55 \\
  $2^-_2$ & $(1p_{1/2})_\Lambda (1d_{3/2})^{-1}_n$  
          & 8.58 & 8.61 & 8.59 & 8.59  & 9.53 \\
  $1^-_2$ & $(1p_{1/2})_\Lambda (1d_{3/2})^{-1}_n$  
          & 8.58 & 8.74 & 8.51 & 8.36  & 9.27 \\
\hline
\end{tabular}
\end{center}
%\bigskip
\caption
{\small
Energies in MeV of the \cala spectrum calculated in MF approximation
and with the three interactions used in our work. Our results are 
compared with the shell model results of Ref. \cite{tze02}. 
We have considered all the states
with energy smaller than 10 MeV. The MF particle-hole transitions
are shown. 
}
\label{tab:40caspectrum}
\end{table}

We observe in Tab. \ref{tab:40caspectrum} that, in our calculations,
the ground state of the \cala system is produced by the
$(1s_{1/2})_\Lambda (1d_{3/2})^{-1}_n$ transition which in MF
approximation gives two degenerate states characterized by angular
momenta and parity 1$^+$ and 2$^+$. All our RPA-like calculations remove
the degeneracy by lowering the 1$^+$ state, while it is the opposite
in the calculation of Ref. \cite{tze02}. In general, the comparison
between our results and those of Ref. \cite{tze02} shows an agreement
in terms of clustering of states around the MF single-particle
transitions. However, within the set of states produced by the same MF
transition, there are some inversions of the multipoles. This happens
for the MF degenerate $2^+_2$ and $3^+$ states, as well as for the set
of degenerate states produced by the $(1p_{3/2})_\Lambda
(1d_{3/2})^{-1}_n$ transition and also for the $(1p_{1/2})_\Lambda
(1d_{3/2})^{-1}_n$ transition. A measure of the experimental spectrum
of \cala would clarify the situation.
%----------------------------------------------------------------------
% 1g9/2 coupling
%----------------------------------------------------------------------
%
\begin{table}[ht]
\begin{center}
\begin{tabular}{c|ccccc}
\hline
\multicolumn{6} {c} {\zrla} \\
\hline
  p   & MF & LN$\delta$ & LND  & LNDE  & exp \cite{hot01a} \\
\hline
 $(1s_{1/2})_\Lambda$ & 0.0 & 0.003 $\pm$ 0.003 & 
                0.005 $\pm$ 0.009  & 0.007 $\pm$ 0.012
                & 0.0 \\ 
 $(1p_{1/2})_\Lambda$ & 4.994 & 5.012 $\pm$ 0.061 & 
                5.010 $\pm$ 0.058 & 4.999 $\pm$ 0.031
                & 4.9 \\ 
 $(1d_{5/2})_\Lambda$ & 10.749 & 10.783 $\pm$ 0.077 & 
               10.766 $\pm$ 0.060 & 10.743 $\pm$ 0.038
                & 11.5 \\ 
 $(1f_{7/2})_\Lambda$ &16.994 & 17.016 $\pm$ 0.035 & 
               17.013 $\pm$ 0.039 & 17.002 $\pm$ 0.029
                & 19.0  \\ 
\hline
\end{tabular}
\end{center}
%\bigskip
\caption
{\small Energies, in MeV, of the various states coupled to the
 $1g_{9/2}$ neutron hole in \zrla, calculated in MF and in our
 RPA-like  
 approach for the three \La-nucleon interactions used in this work.  
 The energies of the RPA-like  
 results are the averages of the various excitation energies
 of each multipole excitation compatible with the indicated single
 particle transition. The experimental energies are those of the 
 $^{89}_\Lambda$Y hypernucleus, and the uncertainty in the 
 determination of the energy is roughly 1 MeV.  
}
\label{tab:1g92}
\end{table}

We have studied the spectra of the \zrla and \pbla hypernuclei and we have
observed that also in these cases the changes with respect to the MF
results are rather small, about 100 keV, of the order of the 2,
maximum 3\% on average. 

As an example of our results, we show in Tab. \ref{tab:1g92} the
energies of the excited states of \zrla obtained by considering the
transitions of the $1g_{9/2}$ neutron hole to the \La single particle
states. This MF picture has been used in the analysis of the
$(\pi^+,K^+)$ reaction on $^{89}_\Lambda$Y target \cite{hot01a}.  In
the same spirit, we have studied the excitation of this neutron hole
into the $s,p,d,f$ states of the \La with $j_\Lambda = l_\Lambda +
1/2$.  We have considered all the possible multipole excitations
compatible with a specific neutron-\La transition.  Obviously, in the
MF calculation all these states have the same excitation energy. Our
RPA-like calculations, however, give different energy values for each
multipole excitation.  In Tab.  \ref{tab:1g92} we compare the MF
energies for each transition with our RPA-like energies obtained by
averaging the results of the various multipole excitations, and we
also show their standard deviation. We observe that, within the
statistical uncertainty, RPA-like and MF results coincide. Also the
agreement with the experimental values is good, especially if we
consider that the data are taken on a $^{89}_\Lambda$Y target, and
they have an uncertainty of 1.65$\pm$0.10 MeV.
%----------------------------------------------------------------------
% 1i13/2 coupling
%----------------------------------------------------------------------
\begin{table}[ht]
\begin{center}
\begin{tabular}{c|cccccc}
\hline
\multicolumn{7} {c} {\pbla} \\
\hline
  p   & MF & LN$\delta$ & LND  & LNDE & exp1 & exp2 \\
\hline
 $(1s_{1/2})_\Lambda$ & 1.633 & 1.630 $\pm$ 0.001 & 
                1.629 $\pm$ 0.002  & 1.623 $\pm$ 0.008
                & 1.6 & -3.0 \\ 
 $(1p_{1/2})_\Lambda$ & 4.621 & 4.629 $\pm$ 0.024 & 
                4.626 $\pm$ 0.025 & 4.616 $\pm$ 0.016
                & 6.8 & 4.6  \\ 
 $(1d_{5/2})_\Lambda$ & 8.220 & 8.237 $\pm$ 0.036 & 
               8.230 $\pm$ 0.030 & 8.216 $\pm$ 0.017
                & 12.84 & 8.24 \\ 
 $(1f_{7/2})_\Lambda$ &12.344 & 12.354$\pm$ 0.020 & 
               12.345 $\pm$ 0.010 & 12.328 $\pm$ 0.014
                & 18.29 & 13.69 \\ 
 $(1g_{9/2})_\Lambda$ &16.926 & 16.948 $\pm$ 0.034 & 
               16.959 $\pm$ 0.040 & 16.926 $\pm$ 0.108
                & 22.10 & 17.49 \\ 
\hline
\end{tabular}
\end{center}
%\bigskip
\caption
{\small Energies, in MeV, of the various states coupled to the
 $1i_{13/2}$ neutron hole in \pbla, calculated in MF and 
 in our RPA-like 
 approach for the three \La-nucleon interactions used in our work. 
 The energies of the RPA-like  
 results are the averages of the various excitation energies
 of each multipole excitation compatible with the indicated single
 particle transition. The values shown in the columns exp1 and exp2
 are the experimental energies of Ref. \cite{has96}, rescaled in two
 different manners (see text). 
}
\label{tab:1i132}
\end{table}

The results of an analogous study for the $1i_{13/2}$ neutron hole in
\pbla coupled with the $s,p,d,f,g$ states of the \La with $j_\Lambda =
l_\Lambda + 1/2$, are shown in Tab. \ref{tab:1i132}.  The RPA-like
energies are all compatible with the MF ones, within the statistical
uncertainty.  In the same table we compare our results with the
energies given in Ref. \cite{has96} where the $(\pi^+,K^+)$ scattering
data on a \pb target have been analyzed in terms of transitions of the
$1i_{13/2}$ neutron hole.  Since, in our calculations, the lowest
energy of \pbla is produced by the transition $(1s_{1/2})_\Lambda
(3p_{1/2})^{-1}_n$, we have positioned the first experimental energy
of Ref. \cite{has96} at 1.63 MeV, which is the difference between the
$3p_{1/2}$ and $1i_{13/2}$ neutron single particle energies. The other
energies have been consistently rescaled, and their values are shown
in Tab. \ref{tab:1i132} by the column labeled exp1.  The agreement is
not satisfactory even if we consider that the experimental energy
resolution is of 2.2 $\pm$ 0.1 MeV.  We then set the position of the
second experimental energy at 4.6 MeV, to match our second state, and
we rescale the other energies.  The corresponding values are shown in
the column exp2. In this case the agreement with our results
noticeably improves.  The source of disagreement with our results is
in the energy difference between the first two states.

\begin{figure}[ht]
\begin{center}
\includegraphics[scale=0.5, angle=0] {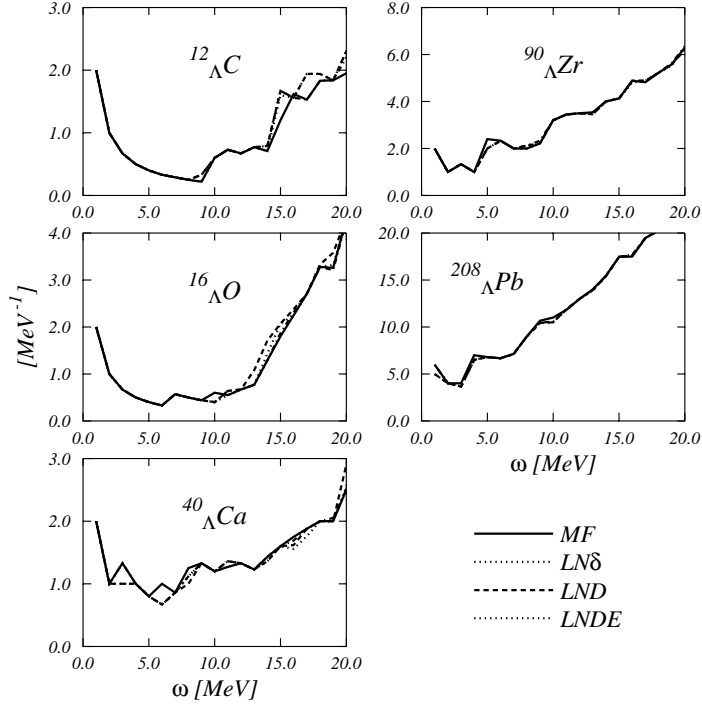} 
\vskip -0.3 cm 
\caption{\small 
  Densities of states as a function of the excitation energy.
  }
\label{fig:dos}
\end{center}
\end{figure}

Up to now, we have presented our results by considering individually
each hypernucleus.  To give a more general view, we show in Fig.
\ref{fig:dos} the density of states of the various hypernuclei as a
function of the excitation energy.  The different lines indicate the
results obtained in MF and RPA-like approaches by using the various
residual interactions.  In our model, the number of excited states for
a given multipolarity depends only on the configuration space, and not
on the residual interaction.  For this reason, differences between the
lines of each panel of the figure indicate the relevance of the
\La-nucleon residual interaction which shifts the excitation energies
with respect to the MF results.  For each hypernucleus, we have taken
into account all the possible multipole excitations which provide a
solution below 20 MeV.
%%%%%%%%%%%%%%%%%%%%%%%%%%%%%%%%%%%%%%%%%%%%%%%%%%%%
% Figure dos
%%%%%%%%%%%%%%%%%%%%%%%%%%%%%%%%%%%%%%%%%%%%%%%%%%%%

A first remark on the results shown in Fig. \ref{fig:dos} is that the
density of states in \pbla is almost ten times larger than that of
\carla or \oxyla. This is an indication of the energy resolution
required to do $\gamma$ spectroscopy in heavy hypernuclei.  A second
remark, more related to the aim of our investigation, is that the
differences between the various lines seem to be more evident for
light hypernuclei than for the heavier ones. This could indicate that
the MF approach becomes more reliable the heavier is the hypernucleus,
or in other words, that collective effects are more inhibited in heavy
hypernuclei than in the light ones.

In order to investigate this issue, we have calculated 
the normalized energy shift with respect to the MF energy,
\begin{equation}
\Delta E = \sum_J^{N_J}  \left(
\frac{1}{N_J}
           \sum_i^S 
\frac {\left|\omega^{J,{\rm RPA}}_i 
                        - \omega^{J,{\rm MF}}_i \right| }
 {\left|\omega^{J,{\rm RPA}}_i 
                        + \omega^{J,{\rm MF}}_i \right| }
\right)
\,\,\,,
\label{eq:delta}
\end{equation}
where we have indicated with $N_J$ the number of multipole 
excitations we have considered, 
and the index $i$ runs on all the $S$ solutions having
the same multipolarity. The above equation is
meaningful since, as we have already mentioned, our model produces the
same number of excited states in both MF and RPA-like calculations,
therefore we can identify the energy shift of each MF solution.  We
show in Tab. \ref{tab:ediff} the $\Delta E$ values obtained for the
various \La-nucleon interactions.  These results indicate that the
deviations from the MF model are less important, the heavier is the
nucleus.
%----------------------------------------------------------------------
% energy differences
%----------------------------------------------------------------------
\begin{table}[ht]
\begin{center}
\begin{tabular}{c|c|ccc}
\hline
       & $J_{\rm max}$ & LN$\delta$ & LND  & LNDE  \\
\hline
 \carla & 5 
     & 0.23  $\pm$  0.07
     & 0.26  $\pm$  0.08
     & 0.22  $\pm$  0.06 \\
 \oxyla & 4 
     & 0.33  $\pm$    0.10
     & 0.30  $\pm$    0.09
     & 0.11  $\pm$    0.03 \\
 \cala & 4  
     & 0.15  $\pm$ 0.04
     & 0.12  $\pm$ 0.04
     & 0.24  $\pm$ 0.07 \\
 \zrla & 6 
     & 0.10   $\pm$   0.02
     & 0.13   $\pm$   0.03
     & 0.09   $\pm$   0.02  \\ 
 \pbla & 10 
     & 0.10   $\pm$   0.02
     & 0.13   $\pm$   0.02
     & 0.09   $\pm$   0.01   \\ 
\hline
\end{tabular}
\end{center}
%\bigskip
\caption
{\small Renormalized energy differences Eq. (\ref{eq:delta}).  We have
also indicated the maximum value of the angular momentum of the
multipole excitation.  }
\label{tab:ediff}
\end{table}

We have compared the values of the hyperonic energy differences of
Tab. \ref{tab:ediff} with analogous values calculated for excitations
of ordinary nuclei. The RPA calculations for the doubly magic nuclei
have been done with the zero-range Landau-Migdal interaction of Ref.
\cite{rin78}. This comparison indicates that collectivity in
hypernuclei is much smaller than in ordinary nuclei.  The $\Delta E$
values for ordinary nuclei are about five times larger than the
analogous values obtained for hypernuclei.

\section{Summary and conclusions}
\label{sec:conclusions}
In this article we have investigated the possible presence of
collective phenomena, beyond the MF model description of hypernuclei.
To this purpose we have developed a particle-hole model where the
formation of the hypernucleus is described as the excitation of a
nucleon into a single particle state describing the \La, and the
hypernucleus state is a linear combination of elementary single
particle excitations, the coefficients of the combinations being
determined by the residual \La-nucleon interaction. Our model is
inspired by the RPA theory and considers some ground state
correlations which can be described in terms of \La-nucleon
excitations. Purely nucleonic correlations, such as those considered
by the traditional RPA theory, are not included.

Our calculations have been done by using a discrete single particle
configuration space, and with three \La-nucleon interactions with
different properties. Since we have been interested in the low energy
part of the spectrum, we have checked the stability of our results
with respect to the size of the configuration space, and therefore 
the scarce relevance of a detailed treatment of the continuum. 

When possible we have compared our results with those obtained by
other calculations.  We obtained a reasonable agreement in \carla and
\oxyla, not only for the energies \cite{ito94,mil01}, but also for the
electromagnetic B-values with shell model results \cite{dal78,mil07}.
For the \cala hypernucleus, the comparison with the shell model
spectrum of Ref.  \cite{tze02} is not so satisfactory.  We have the
same clustering of states in the spectrum, but we often obtain an
inversion of states within the cluster.

Concerning the comparison with the experimental data, we obtain a
reasonable agreement with the energies measured in \oxyla.  Our model
is unable to predict two states observed in \carla because it does not
consider the partial occupation of the $1p_{1/2}$ neutron state.  The
energy resolution of \cala data \cite{bru78,pil91} does not allow a
meaningful comparison with our results. For the \zrla and \pbla
hypernuclei we made a comparison with data analyzed in terms of
excitation of a single neutron \cite{hot01a,has96}. The agreement with
\zrla data is satisfactory, while for the \pbla case there are some
problems regarding the spacing between the first two states.

To investigate the presence of collective effects in the spectra of
hypernuclei, we have calculated the form factors of the transition
producing the hypernucleus and we have compared MF and our RPA-like
results.  In few cases we have found remarkable differences, however,
in all the cases but one, these differences could be explained by
considering the excited state as combination of only two single
particle transitions.  This is not really what we would call a
collective behavior of the system. The only situation standing out of
this trend is that of the 0$^+$ excitation in \oxyla around 10 MeV,
which seems to have a more collective structure.

The energy shifts with respect to the MF results are indicators of
effects coming from the \La-nucleon interaction, and possibly of
collective phenomena. Densities of states and average energy shifts do
not show remarkable features beyond the MF approach. In our
calculations the differences between MF and our RPA-like results
become smaller the heavier is the hypernucleus. This is an indication
that perturbation effects produced by the \La on the nuclear systems
are better averaged in a heavy system.

The average energy shifts for hypernuclear excitations are, at most, 
20\% of the shifts obtained for the excitation of a purely
nucleonic system.  This result is almost independent of the number of
nucleons involved, and indicates that in hypernuclear excitation,
effects beyond the MF are much less important than in the excitation
of the ordinary nuclear system.

Our model can be further improved in several ways.
One of them consists in
performing self-consistent calculations, where the single particle
basis is generated in a Hartree-Fock approach, with the same
\La-nucleon interaction used in the RPA-like calculations.  Other
improvements can be 
obtained by considering nucleonic correlations in the
ground state. This could be done in terms of hole-particle
excitations, as in the ordinary RPA, in terms of partial occupation
probability of the nucleonic single particle states, as it happens in
the quasi-particle RPA, and also by considering explicitly the
short-range correlations as in Ref. \cite{ari01}.

\vskip 2.0 cm
{\bf ACKNOWLEDGMENTS} \\ 
We thank B. Dalena for useful discussions. 
\vskip 3.0 cm

%%%%%%%%%%%%%%%%%%%%%%%%%%%%%%%%%%%%%%%%%%%%%%%%%%%%%%%%%%%%%%%%%%%%%%%
\section*{Appendix}
In this appendix we give the expressions of the matrix elements 
(\ref{eq:vj}) and (\ref{eq:uj}) of the secular RPA-like equations  
for all the terms of the interaction (\ref{eq:fourier}).

For the scalar term $V^c_{\Lambda N}$ we obtain the expressions 
\begin{eqnarray}
\nonumber 
v_{c,\Lambda h\Lambda'h'}^{J,\,dir}
&=& \frac{2}{\pi} \int dqq^2\,F_\Lambda (q) 
     \int dr_1r^2_1R_\Lambda^\ast(r_1)
    R_h(r_1)j_J(qr_1) \\
\nonumber &~&
 \int dr_2r^2_2 R_{h'}^\ast (r_2)R_{\Lambda'}(r_2)j_J(qr_2)  \\
\nonumber &~&
\frac{(-1)^{j_\Lambda+j_{\Lambda'}+1}}{4\pi}
\hat{j}_\Lambda \hat{j}_h \hat{j}_{\Lambda'} \hat{j}_{h'}
\threej {j_\Lambda} {J}{ j_h} {\half} {0} {-\half}
\threej {j_{h'}} {J}{ j_{\Lambda'}} {\half} {0} {-\half} \\
&~&
\xi(l_\Lambda+l_h+J)\xi(l_{\Lambda'}+l_{h'}+J)
\,\,,
\label{eq:v1d}
\end{eqnarray}
and
\begin{eqnarray}
\nonumber 
v_{c,\Lambda h\Lambda'h'}^{J,\,exch}&=& \frac{2}{\pi} 
\sum_{l}\int  dqq^2 \,F_\Lambda (q) 
\int dr_1r^2_1R_\Lambda^\ast(r_1)
R_{\Lambda'}(r_1)j_l(qr_1) \\
\nonumber &~&
\int{dr_2r^2_2R_{h'}^\ast (r_2)R_h(r_2)j_l(qr_2)} \\
\nonumber &~&
\frac{(-1)^{j_\Lambda+j_{\Lambda'}+J+l}}{4\pi}
\hat{j}_\Lambda \hat{j}_h \hat{j}_{\Lambda'} \hat{j}_{h'} \hat{l}^2
\threej {j_\Lambda} {l}{ j_{\Lambda'}} {\half} {0} {-\half}
\threej {j_{h'}} {l}{ j_{h}} {\half} {0} {-\half}
\nonumber\\
&& \sixj {j_\Lambda} {j_h}{J} {j_{h'}} {j_{\Lambda'}} {l}
\xi(l_\Lambda+l_{\Lambda'}+l)\xi(l_h+l_{h'}+l) 
\label{eq:v1e}
\,\,.
\end{eqnarray}
In the above equations we have indicated with $R(r)$ the radial part
of the single particle wave functions, with $j_l(qr)$ the spherical
Bessel functions, and with the traditional symbols the Wigner three
and six $j$ coefficients \cite{edm57}. 

The analogous expressions for the spin-dependent terms are 
\begin{eqnarray}
\nonumber 
v_{\sigma,\Lambda h\Lambda'h'}^{J,\,dir} &=& \frac{2}{\pi}
\sum_l\int dqq^2 \,  G_\Lambda(q) 
\int dr_1r^2_1R_\Lambda^\ast(r_1) R_h(r_1)j_l(qr_1) \\
\nonumber &~&
\int dr_2r^2_2R_{h'}^\ast (r_2) R_{\Lambda'}(r_2)j_l(qr_2) \\
\nonumber &~&
(-1)^{l+J+l_\Lambda+l_{h'}+j_{\Lambda'}+j_{h'}}
\frac{3}{2\pi}
\hat{l}^2 \hat{l}_\Lambda \hat{l}_h \hat{l}_{\Lambda'}
\hat{l}_{h'} \hat{j}_\Lambda \hat{j}_h \hat{j}_{\Lambda'} \hat{j}_{h'}
\nonumber\\
&& \ninej {l_\Lambda} {\half}{j_\Lambda}
          {l_h} {\half}{j_h}
          {l}   {1} {J}
   \ninej {l_\Lambda'} {\half}{j_\Lambda'}
          {l_h'} {\half}{j_h'}
          {l}   {1} {J}
   \threej {l_\Lambda}{l}{l_h}{0}{0}{0}    
   \threej {l_\Lambda'}{l}{l_{h'}}{0}{0}{0}
\,\,,
\label{eq:v3d}
\end{eqnarray}
and
\begin{eqnarray}
\nonumber 
v_{\sigma,\Lambda h\Lambda'h'}^{J,\,exch} &=& \frac{2}{\pi} 
\sum_l\int dqq^2 \, G_\Lambda(q) 
\int dr_1r^2_1R_\Lambda^\ast(r_1) R_{\Lambda'}(r_1)j_l(qr_1) \\
&~& \nonumber
\int dr_2r^2_2R_{h'}^\ast (r_2)R_{h}(r_2)j_l(qr_2) \\
\nonumber
&~&\sum_{L}\frac{3}{2\pi}(-1)^{l_\Lambda+l_{h'}+j_{\Lambda'}+j_{h'}+l+J+1}
\\ 
\nonumber &~&
\hat{l}_\Lambda \hat{l}_h \hat{l}_{\Lambda'} \hat{l}_{h'} 
\hat{j}_\Lambda \hat{j}_h \hat{j}_{\Lambda'}
\hat{j}_{h'} \hat{l}^2 \hat{L}^2
\sixj {j_\Lambda} {j_h} {J} {j_{h'}} {j_{\Lambda'}} {L}
\nonumber\\
&& \ninej {l_\Lambda} {\half}{j_\Lambda}
          {l_h} {\half}{j_h}
          {l}   {1} {L}
   \ninej {l_\Lambda'} {\half}{j_\Lambda'}
          {l_h'} {\half}{j_h'}
          {l}   {1} {L}
   \threej {l_\Lambda}{l}{l_{\Lambda'}}{0}{0}{0}    
   \threej {l_h}{l}{l_{h'}}{0}{0}{0}
\,\,.
\label{eq:v3e}
\end{eqnarray}

For the tensor components of the interaction we obtain:
\begin{eqnarray}
v_{t,\Lambda h\Lambda'h'}^{J,\, dir} &=&
\frac{2}{\pi}
\sum_{l_1l_2}
\int dqq^2 \,  H_\Lambda(q) 
\int dr_1r^2_1 R_\Lambda^\ast(r_1) R_{h}(r_1)j_{l_1}(qr_1)  
\nonumber \\
&&\int dr_2r^2_2R_{h'}^\ast (r_2)R_{\Lambda'}(r_2)j_{l_2}(qr_2) 
\nonumber \\
&&\frac{3\sqrt{30}}{2\pi}(-1)^{j_{\Lambda'}+j_{h'}+l_{\Lambda}+l_{h'}}
\hat{l}_\Lambda \hat{l}_h \hat{l}_{\Lambda'} \hat{l}_{h'}
\hat{j}_\Lambda \hat{j}_h \hat{j}_{\Lambda'} \hat{j_{h'}} 
\nonumber\\
&&
(-1)^{\frac{3}{2}(l_1+l_2)}\hat{l_1}^2\hat{l_2}^2
 \threej {l_1}{l_2}{2}{0}{0}{0}
 \threej {l_\Lambda}{l_1}{l_h}{0}{0}{0}
 \threej {l_{h'}}{l_2}{l_{\Lambda'}}{0}{0}{0}
  \nonumber\\
&& \sixj {l_1}{l_2}{2}{1}{1}{J}
   \ninej {l_\Lambda}{\half}{j_\Lambda}
          {l_h}{\half}{j_h}
          {l_1}{1}{J}
   \ninej {l_{h'}}{\half}{j_{h'}}
          {l_{\Lambda'}}{\half}{j_{\Lambda'}}
          {l_2}{1}{J}
\,\,,
\label{eq:v5d}
\end{eqnarray}
for the direct term, and for the exchange term we get the expression 
\begin{eqnarray}
v_{t,\Lambda h\Lambda'h'}^{J,\,exch} &=&
- \frac{2}{\pi} \sum_{l_1l_2}\int dqq^2 \, H_\Lambda(q) 
\int dr_1r^2_1R_\Lambda^\ast(r_1) R_{\Lambda'}(r_1)j_{l_1}(qr_1) 
\nonumber \\
&& \int dr_2r^2_2R_{h'}^\ast (r_2)R_{h}(r_2)j_{l_2}(qr_2) 
\nonumber\\
\nonumber
&&\frac{3\sqrt{30}}{\pi^2}
(-1)^{l_\Lambda+l_{h'}+j_{\Lambda'}+j_{h'}+l_1+l_2+J}i^{l_1+l_2} \\
&~&
\hat{l}_\Lambda \hat{l}_h \hat{l}_{\Lambda'} \hat{l}_{h'}
\hat{j}_\Lambda \hat{j}_h \hat{j}_{\Lambda'} \hat{j}_{h'}
\hat{l}_1^2 \hat{l}_2^2
\nonumber\\
&& \threej {l_1}{l_2}{2}{0}{0}{0}
 \threej {l_\Lambda}{l_1}{l_{\Lambda'}}{0}{0}{0}
 \threej {l_{h'}}{l_2}{l_{\Lambda}}{0}{0}{0}
\nonumber\\
&&\sum_L(-1)^L\hat{L}^2
\sixj {l_1}{l_2}{2}{1}{1}{L} 
\sixj {j_\Lambda}{j_h}{J}{j_{h'}}{j_{\Lambda'}}{L}
\nonumber\\
&&
   \ninej {l_\Lambda}{\half}{j_\Lambda}
          {l_h}{\half}{j_h}
          {l_1}{1}{L}
   \ninej {l_{h'}}{\half}{j_{h'}}
          {l_{\Lambda'}}{\half}{j_{\Lambda'}}
          {l_2}{1}{L}
\,\,.
\label{eq:v5e}
\end{eqnarray}

%\newpage
%\bibliographystyle{elsart-num} 
%\bibliography{miei,fhnc,em,altro,rpa,tesi,libri,hyp,nuovi}

%
%
\end{document}